# Motion In-Betweening with Phase Manifolds

PAUL STARKE, Electronic Arts, Universität Hamburg
SEBASTIAN STARKE, Meta Reality Labs
TAKU KOMURA, The University of Hong Kong
FRANK STEINICKE, Universität Hamburg

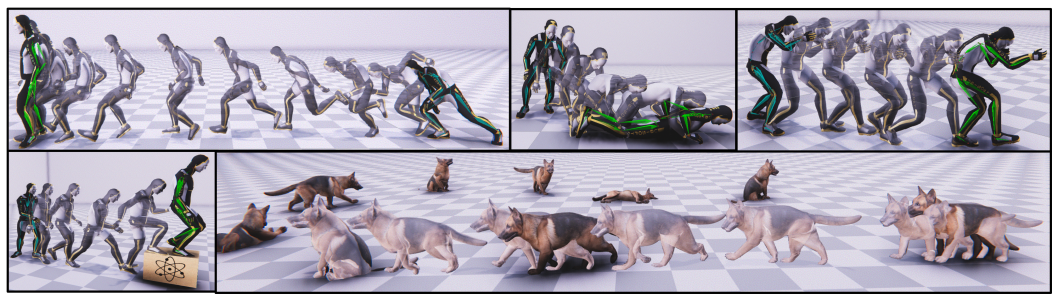

Fig. 1. A selection of motion transitions between a start keyframe and a target keyframe as synthesized by our system: Given a target frame and a desired transition duration, our model can synthesize motion types to reach the target in a natural manner.

This paper introduces a novel data-driven motion in-betweening system to reach target poses of characters by making use of phases variables learned by a Periodic Autoencoder. Our approach utilizes a mixture-of-experts neural network model, in which the phases cluster movements in both space and time with different expert weights. Each generated set of weights then produces a sequence of poses in an autoregressive manner between the current and target state of the character. In addition, to satisfy poses which are manually modified by the animators or where certain end effectors serve as constraints to be reached by the animation, a learned bi-directional control scheme is implemented to satisfy such constraints. The results demonstrate that using phases for motion in-betweening tasks sharpen the interpolated movements, and furthermore stabilizes the learning process. Moreover, using phases for motion in-betweening tasks can also synthesize more challenging movements beyond locomotion behaviors. Additionally, style control is enabled between given target keyframes. Our proposed framework can compete with popular state-of-the-art methods for motion in-betweening in terms of motion quality and generalization, especially in the existence of long transition durations. Our framework contributes to faster prototyping workflows for creating animated character sequences, which is of enormous interest for the game and film industry.

Authors' addresses: Paul Starke, Electronic Arts, Universität Hamburg, paulstarke.ps@gmail.com; Sebastian Starke, Meta Reality Labs, sstarke@ea.com; Taku Komura, The University of Hong Kong, taku@cs.hku.hk; Frank Steinicke, Universität Hamburg, frank.steinicke@uni-hamburg.de.

## 1 INTRODUCTION

Data-driven motion in-betweening is attracting the interest of researchers and engineers in the computer graphics domain due to its simplicity for synthesizing motion and its workflow, which is highly compatible with traditional keyframe animation. In combination with interactive animation synthesis tools, animators can produce realistic character animations from a small number of keyframes either manually designed or imported from a motion capture dataset. Despite the simplicity and effectiveness of motion in-betweening, existing methods can suffer from poor quality in the synthesized motions, especially when the duration between the keyframes is long and inconsistent. In such a situation, the system tends to simply linearly interpolate the keyframes, making the motion appear unrealistic and smoothed out. They also tend to perform poorly when the number of training data is small, which reveals their low generalizability to unseen keyframes during runtime.

To address the limitations of previous methods, we propose a novel motion in-betweening framework, which makes use of a phase prior learned from a motion capture database. By considering the periodicity of the movements, the system can synthesize a correct motion even for keyframes, which have long time gaps. Furthermore, the system can also produce realistic movements and transitions for actions that are rarely available in datasets, such as backward walking and side-stepping. We conducted various experiments to show the robustness and versatility of our approach. Compared to existing state-of-the-art motion in-betweening techniques, our system can produce better results both qualitatively and quantitatively. We also evaluate the method through various stress tests and a comprehensive ablation study.

The key contribution of this paper is in the engineering of a state-of-the-art motion in-betweening system that combines the latest advances in character animation research which allows for:

- A robust motion in-betweening scheme that can animate realistic movements in real-time using phase variables and
- capability of handling longer transition times, spatial large target distances, and unseen target postures that is examined through a comprehensive evaluation.

We have made the source code[1] for our system publicly available.

## 2 RELATED WORK

*Keyframes as an Interface for Motion Synthesis.* Keyframe animation is a traditional approach for synthesizing character animation. Since many keyframes are required for producing high-quality animations, alternative techniques, which make use of physics and motion capture data, have been proposed. Spacetime constraints [Witkin and Kass 1988] minimize an energy based on momentum and use the keyframes as constraints that the character should pass through. Liu and Popovic [Liu and Popović 2002] produce realistic human motion such as jumping and flipping using spacetime constraints. Similarly, Fang and Pollard [2003] interpolate keyframes using an energy function based on momentum. Spacetime constraints purely based on keyframes and momentum optimization have difficulty producing low energy motions such as ordinary walking due to the other factors such as biomechanics that play a heavy role for producing realistic human movements. As such, later the focus shifted to make use of motion capture data.

---

[1]https://github.com/pauzii/PhaseBetweener

Keyframes can also function as a good interface for editing motion capture data [Gleicher 1997; Popović and Witkin 1999; Witkin and Popovic 1995]. Witkin and Popovic [1995] propose to fade-in/out offsets with keyframes for motion editing. Gleicher [1997] applies spacetime optimization for such edits. Popovic and Witkin [1999] apply the concept of spacetime constraints to human body models for physically editing motion capture data. PD targets of physics-based animation can also be considered keyframes for character motion control. Such PD controllers have been applied for animating locomotion [Hodgins and Pollard 1997; Yin et al. 2007], athletic movements [Hodgins et al. 1995], and contact-rich movements [Liu et al. 2010]. Recently, reinforcement learning controllers have been applied for optimizing PD targets to follow motion capture data [Bergamin et al. 2019; Peng et al. 2018, 2017; Won and Lee 2019]. Besides, Gopinath et al. [2022] conduct motion in-betweening in the physics domain.

*Data-Driven Motion Synthesis with Keyframes.* As more motion capture data has become available, the focus of research has switched to making use of the motion database for synthesizing movements, instead of editing short motion clips. Arikan and Forsyth [2002] use a hierarchy of graphs to represent connectivity of a motion database and perform randomized search to identify motions that satisfy user constraints given by keyframes. Safonova et al. [2004] conduct Principal Component Analysis (PCA) to the motion capture data and use them for interpolating keyframes. Sung et al. [2005] interpolate poses in far distance by traversing motion graphs. Kovar et al. [2004] provide an IK-interface for interpolating motions of k-nearest neighbors. Mukai and Kuriyama [2005] similarly present such an interface for interpolating motions by Krigging. Min and Chai [2012] provide a similar interface for editing the motions synthesized by functional PCA. Classic machine learning methods have limitations in terms of nonlinearity or scalability, that hamper their application for complex and long movements.

*Neural Motion In-betweening.* The advancement of neural networks has made it possible to learn the human motion manifold [Holden et al. 2015; Wang et al. 2019], facilitating the synthesis of new motion data using diverse control signals. The pioneering research by Harvey et al. [2020] demonstrates that neural networks are capable of creating plausible interpolation movements for a given pair of keyframe poses. Their method's key contribution is a time-embedding, which is intended to strengthen the correlation between timing and motion. Tang et al. [2022] extend these findings with a Convolutional Variational Autoencoder (CVAE). Duan et al. [2022] introduce a motion perceptual loss, which is defined by the context of motion, and, hence, focuses on certain features of motions. Previous motion in-betweening works often suffer from generalizing the timing information, e.g., difficulty in adapting to longer duration between keyframes, or interpolating poses with different phases of locomotion. Such timing information has first been successfully demonstrated with Phase-Functioned Neural Networks (PFNN) [Holden et al. 2017], where a phase controls the interpolation of human locomotion behaviors for neural character controllers. Building on this approach, Starke et al. [2019] demonstrate how different motion skills could be represented by multiple phase variables that would then enable the model to learn a large range of movements in high quality. Subsequently, this representation of phase variables has first been formulated on a local joint level [Starke et al. 2020] for contact-based movements, and most recently in a learning-based fashion [Starke et al. 2022] to extract the spatial-temporal alignment of arbitrary character movements. However, those methods have been mainly demonstrated for interactive character control tasks with fixed time window representations, and are in our work further explored for motion interpolation tasks with dynamic time intervals between control signals, such as keyframe poses.

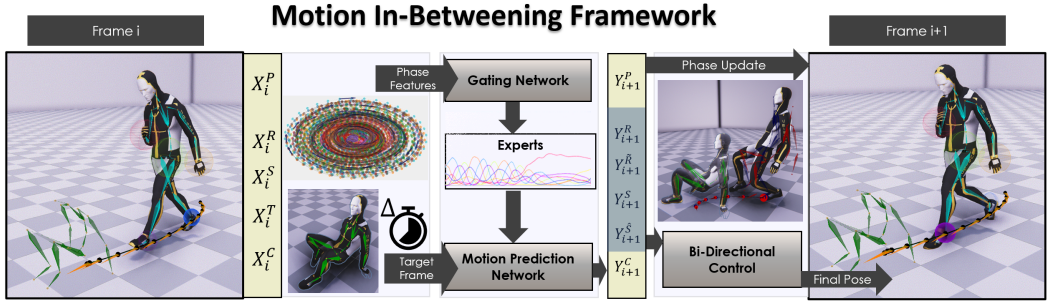

Fig. 2. The architecture of the system is composed of the gating network and the motion prediction network. The gating network takes as input the phase segments that are learned by the Periodic Autoencoder and computes the blending coefficients to generate the motion prediction network. The motion prediction network takes as input the current pose, trajectory, contacts and the control variables of the target frame to predict the motion in the next frame.

## 3 MOTION IN-BETWEENING FRAMEWORK

Our motion in-betweening framework is a time-series system that predicts the character pose from one frame into the next in an autoregressive fashion. The system utilizes a mixture-of-experts scheme similar to [Zhang et al. 2018] and [Starke et al. 2019] consisting of the motion prediction network and the gating network. We train the model in an end-to-end manner using extracted phase information and pose data from the motion capture. We implement two key techniques in our framework to generate the in-betweened motions, which are phase manifolds – to achieve a better temporal coherence and detail of the motion – and a bi-directional control scheme – to improve the accuracy for reaching the target pose.

### 3.1 Applying Phase Manifolds

A key challenge for generating movements between two given target poses is to produce a motion that progresses realistically in a smooth and temporally coherent manner. This is possible to achieve for smaller time windows, where the possible set of transitions is limited, i.e. 1-2 seconds, but much more challenging when the time windows between the key poses becomes larger, i.e. 4 seconds and beyond. One reason is that for a same target pose to be reached in the future, there can be multiple poses at a certain timestamp before that, which leads to inherent ambiguity and a significantly growing set of transitions for larger in-betweening times. This leads to the problem that poses of different timestamps become blended and may be difficult to smoothly progress forward in time. In contrast, phase features intrinsically encode the motion in the time domain which helps alleviate two main challenges: First, at every updated frame during runtime, they can ensure that only poses from a valid set of transitions from the next timestamp are sampled by the model. Second, the motion is represented by a manifold in the frequency domain that can reproduce motion in a longer horizon beyond the transition duration. Integrating phase updates predicted by the model in form of frequencies, the model can regress movements beyond the used time horizons during training. To obtain those phase features from our dataset, we utilize the Periodic Autoencoder in [Starke et al. 2022] and train our in-betweening model to use them as input and predict the phase updates in the output. The phase manifold $\mathcal{P}_i$ for each input window around frame $i$ is computed from the extracted vectors of amplitudes $A_i$ and phases $\Theta_i$:

$$\mathcal{P}_i = A_i \cdot \begin{pmatrix} sin\Theta_i \\ cos\Theta_i \end{pmatrix}. \tag{1}$$

This helps the model to achieve a better temporal coherence while approaching the target pose as well as a significantly better generalization to produce movements for much longer time windows than seen during training.

## 3.2 Bi-directional Control

When generating movements in-between keyframes, guaranteeing that a pose accurately terminates at a certain target location can be difficult to achieve when using neural networks. This can happen due to error accumulation of pose updates over multiple frames or smoothing of movements due to compression by the network. To alleviate such problems we utilize a bi-directional control scheme similar to [Starke et al. 2019] in which the motion is inferred from both the ego-centric and goal-centric point of view. In contrast to [Starke et al. 2019] which only predicts the root trajectory in both spaces to navigate the character, we additionally predict all joint transformations of the entire character motion in both spaces. The trajectory and character pose predictions are then blended during runtime and fed back into the neural network in the next frame to increase the precision of the character to reach the desired pose (see Fig. 9). The predictions in both ego-centric and goal-centric spaces are blended as

$$(1-\lambda)Y_{i+1}^S + \lambda T_{i+1}Y_{i+1}^{\hat{S}}, \tag{2}$$

where $\lambda$ is the blending parameter, $T_{i+1}$ is the goal-to-ego-centric transformation matrix, and $Y^S$ and $Y^{\hat{S}}$ are the goal- and ego-centric predictions for the character state and root trajectory. The blending parameter is computed using a smooth-step activation function based on the remaining time until reaching the target pose, which smoothly drives the character towards the target pose in the desired in-betweening time.

## 4 NETWORK TRAINING

This section presents the training stage of the neural network and describes data processing and input/output features of the system. We use the full LaFAN1 dataset [Harvey et al. 2020] from Ubisoft to train our model. Similar to [Harvey et al. 2020] and [Tang et al. 2022], we use the Subject 5 captures with 13 animation sequences as the test set and from the Subject 1 to 4 we extract features for training the network. In total, 64 sequences are processed that contain 399,139 motion frames (∼3.69 hours at 30fps). The final training data is augmented by mirroring. Our network model is a Mixture-of-Experts system similar to [Zhang et al. 2018], which contains 3 fully-connected layers that are blended by a set of 8 expert weights. Those experts are interpolated by predicting their blending parameters from a gating network using the phase features. The network is trained end-to-end and using MSE loss across all features.

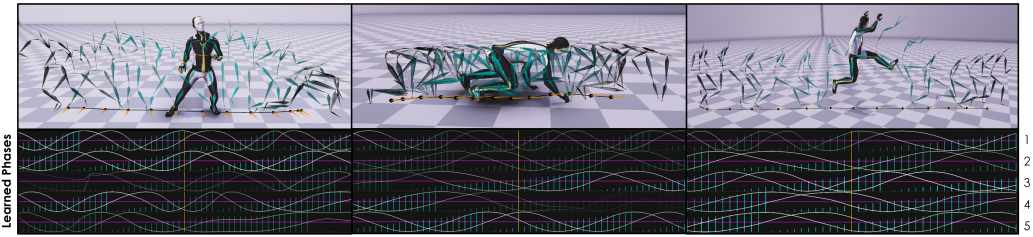

Fig. 3. Extracted phase channels of a two second time window containing sidestepping (left), crawling (middle), and run to jump (right) motion behaviors. The bottom plots represent the phase channels extracted with the Periodic Autoencoder. The white curve defines the 1D phase value and the rotating 2D phases are visualized by the blue bars with the amplitudes represented by the opacity of the bars.

## 4.1 Feature Extraction

Using the skeletal character motion, we extract the root trajectory and contact labels from the captured data. The root trajectory is computed similar to [Holden et al. 2017] and the contacts are computed similar to [Starke et al. 2020] using thresholds for feet height and velocity. Additionally, for each input frame we sample a future pose at frame $i + \Delta t$ where $\Delta t$ is a uniform random time offset between $\frac{1}{30}$ and 2 seconds (1 to 60 frames) to the target. This generates a distribution of different in-betweening duration that can occur during inference for a current character state. We further learn phase features from the data as a pre-processing stage using the Periodic Autoencoder framework in [Starke et al. 2022], which aligns all movements across the entire motion dataset in an unsupervised manner. Examples for extracted phases are visualized in Fig. 3, which shows that different phase alignments are extracted for different character motions. The Periodic Autoencoder is trained by using the 3D velocities of all joint trajectories relative to the root space of the character within a 2 seconds window. Training the phase network takes ~3 hours for 10 epochs. The learned phase features are added to the existing motion dataset without the need of retaining the phase network. More details about the input and output features will be discussed in the following section.

## 4.2 Input and Output Formats

The system is a time-series model that predicts the state variables of the character, future trajectory, contacts, and phase updates in the next frame $i + 1$ given those in the current frame $i$. Some features live in a time-series window $\mathcal{T}_{-1s}^{1s}$ within which data of 13 uniformly-sampled points (6 for each 1s future and past window, and one for the current frame) are collected. The notation $\mathcal{T}_{t_0}^{t_1} = N$ describes the collection of $N$ data samples within a time window of $t_0 \leq t \leq t_1$.

*Inputs.* The complete input vector $X_i$ at frame $i$ consists of five components $X_i = \{X_i^{\mathcal{R}}, X_i^{\mathcal{S}}, X_i^{\mathcal{T}}, X_i^{C}, X_i^{\mathcal{P}}\}$ where each item is described below.

- **Past/Future Root Trajectory in Target Coordinate System** $X_i^{\mathcal{R}} = \{T_i^{\hat{p}}, T_i^{\hat{r}}, T_i^{\hat{v}}, T_i^{\Delta}\}$ consists of the following variables that are sampled in the past-to-current time window $\mathcal{T}_{-1s}^{1s} = 13$. For controlling the characters locomotion the system is trained in the 2D-horizontal plane relative to the target root coordinate frame $i$ of trajectory positions $T_i^{\hat{p}} \in \mathbb{R}^{2\mathcal{T}}$, forward facing trajectory directions $T_i^{\hat{r}} \in \mathbb{R}^{2\mathcal{T}}$ and trajectory velocities $T_i^{\hat{v}} \in \mathbb{R}^{2\mathcal{T}}$. A time delta offset $T_i^{\Delta} \in \mathbb{R}^{\mathcal{T}}$ to the current target pose is embedded at each sampled point of the trajectory.
- **Character State** $X_i^{\mathcal{S}} = \{p_i, r_i, v_i\}$ represents the state of the character with $\mathbb{B} = 22$ bones at the current frame $i$. It consists of the bone positions $p_i \in \mathbb{R}^{3B}$, bone rotations $r_i \in \mathbb{R}^{6B}$, and bone velocities $v_i \in \mathbb{R}^{3B}$, where each bone rotation is formulated by its pair of Cartesian forward and up vectors to create an unambiguous and continuous interpolation space. For training, these Cartesian features are transformed into the root coordinate frame of the character at frame $i$.
- **Target State** $X_i^{\mathcal{T}} = \{\tilde{p}_i, \tilde{r}_i\}$ are the variables that describe the state of the target frame with respect to the character in frame $i$. Those consist of bone positions $\tilde{p}_i \in \mathbb{R}^{3B}$ and bone rotations $\tilde{r}_i \in \mathbb{R}^{6B}$ with $\mathbb{B} = 22$ bones.
- **Contact Information** $X_i^{C} = \{C_i\}$ conditions the generated motion and are sampled along the past-to-current time series window $\mathcal{T}_{-1s}^{0s} = 7$. The contacts $C_i \in \mathbb{R}^{5\mathcal{T}}$ for feet, hands, and hip that appeared during the past are stabilizing the movements.
- **Motion Phases** $X_i^{\mathcal{P}} = \{\Theta_i \in \mathbb{R}^{2C\mathcal{T}}\}$ are each represented by 2D phase vectors of changing amplitude that are sampled along the past to future time series window $\mathcal{T}_{-1s}^{1s} = 13$. The number of extracted phases is $C = 5$.

The learned phases $X^{\mathcal{P}}$ are fed into the gating network, and the rest is fed into the motion prediction network.

*Outputs.* The output vector $Y_{i+1} = \{Y_{i+1}^R, Y_{i+1}^{\hat{R}}, Y_{i+1}^S, Y_{i+1}^{\hat{S}}, Y_{i+1}^C, Y_{i+1}^P\}$ for the next frame $i + 1$ is computed by the motion prediction network, and consists of the following components.

- **Future Root Trajectory** $Y_{i+1}^R = \{T_{i+1}^p, T_{i+1}^r, T_{i+1}^v\}$ are the future (control) signals each sampled along the current to future time series window $\mathcal{T}_{0s}^{1s} = 7$ of the next frame $i+1$ and are relative to the updated root coordinate system of the character in that frame.
- **Future Root Trajectory in Target Coordinate System** $Y_{i+1}^{\hat{R}} = \{T_{i+1}^{\hat{p}}, T_{i+1}^{\hat{r}}, T_{i+1}^{\hat{v}}\}$ are the same features as above but are defined relative to the target postures root coordinate frame.
- **Predicted Character Pose** $Y_{i+1}^S = \{p_{i+1}, r_{i+1}, v_{i+1}\}$ is the predicted pose and joint velocities of the character at next frame $i + 1$ with $\mathbb{B} = 22$ bones relative to the root coordinate system.
- **Predicted Character Pose in the Target Joint Coordinate System** $Y_{i+1}^{\hat{S}} = \{\hat{p}_{i+1}, \hat{r}_{i+1}, \hat{v}_{i+1}\}$ are the predicted bone positions, rotations and velocities of the character at next frame $i + 1$ but are transformed into semi-joint coordinate systems of the target pose. These coordinate systems represent the target joint position but the rotation is defined by the target poses root coordinate system.
- **Predicted Contacts** $Y_{i+1}^C$ with $c_{i+1} \in \mathbb{R}^5$ are the key joint contact labels where the key joints are feet, hands and hip.
- **Motion Phase Updates** $Y_{i+1}^P, Y_{i+1}^{P_F}, Y_{i+1}^{P_A}$ are predicted for the future time series samples $\mathcal{T}_{0s}^{1s} = 7$ of the next frame, where $P_F$ and $P_A$ are the phase frequency and amplitudes to update the current phase and interpolate with the next phase state similar to [Starke et al. 2022].

## 5 EXPERIMENTS AND EVALUATION

The experiments are conducted on an Alienware x17 R2 laptop with Intel i9-12900HK processing cores and a NVIDIA GeForce RTX 3080Ti GPU, requiring 7-9ms per frame for each character including control parameter processing, inference time and scene rendering. The animation is run at 30Hz framerate and the framework is implemented in the Unity 3D engine. The system allows setting the target pose and in-betweening time as control parameters. The poses can be given either by sampling from the motion capture animation clips or by the user (see Fig. 10). The authoring system then automatically computes a smooth path between them, which can additionally be modified by the user, such as for setting a desired facing direction of the character. An example of the workflow and rendered results can be seen in the accompanying video. The movements are generated in a fully autoregressive fashion from the fixed target pose and the given in-betweening duration. We utilize predicted contact labels to apply an inverse kinematics postprocessing on the feet end-effectors to reduce remaining foot skating artefacts.

The evaluation focuses on the motion quality, transition quality and model generalizability under unseen control signals, particularly extrapolation to longer time windows than observed during training. The readers are referred to the supplementary video for animated results.

### 5.1 Impact of Phase Variables

First, we evaluate the impact of phase variables for the motion in-betweening system. The system is demonstrated using learned phase manifolds [Starke et al. 2022] with contact-based phases [Starke et al. 2020] and compared to a standard neural network where the phase is not given as an input. Figure 4 shows generated motion sequences with a transition time of 2 seconds. When using the network that does not utilize phase inputs, the system is blending movements of different timings which leads to partly stiff and unnatural movements, particularly for non-locomotion behaviors. Using the learned phases as input, the difference is most obvious when viewing very agile motions, such as crawling, where the phase manifold enables better transitions. Compared to contact-based phases, the learned phases enable more detailed motion as the features contain more information that align the upper body as well as the lower body movements. This is especially the case for motions that do not produce contacts for certain limbs, and contact-based phases would then cover no information for those channels. In general, the learned phases prevent the model from producing excessive drifting or smoothing artefacts: Using phases, the movements are intrinsically aligned in time where each phase maps to a certain set of transitions, which is not the case when solely relying on the remaining in-betweening time to reach the target where same time states can map to multiple and very different poses.

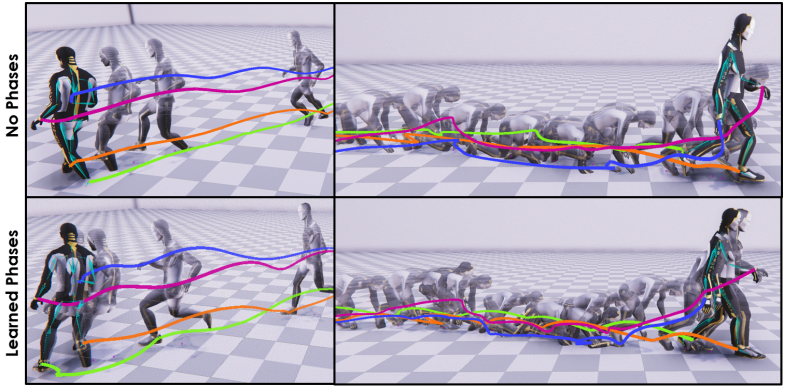

Fig. 4. The motion transitions synthesized by the network with/without using phase features as inputs and control parameters for each model. The left side shows a fast sprint behavior with a very sharp turn in the end and the control parameters on the right side are set to generate a crawl to stand-up motion.

We evaluate the movements in terms of vividness and foot skating artefacts. For evaluating vividness, the average joint rotations per second for different motion skills are computed and shown in Table 1 for a wide range of motion types, and Figure 5 and 6 provide an evaluation on different locomotion classes specifically. Since motions synthesized by neural networks tend to smooth the motion, this metric provides a good indication of how agile the generated motions appear. Blurry motion artefacts are heavily reduced by using the phase features. The feet contact accuracy is evaluated by computing their horizontal movement when the height and vertical velocity are below certain thresholds (1.5cm and 1m/s). Fig. 6 shows that the model with learned phase variables produces the least foot skating artefacts.

Table 1. The average joint rotations per second for different classes of motion transitions. The control variables are extracted from the test data with a transition duration time between 1 and 3 seconds.

| Transition | No Phases | Contact Phases | Learned Phases | GT |
|---|---|---|---|---|
| Walk | 55.2 | 61.7 | **63.4** | 66.4 |
| Sprint with Drastic Turn | 96.9 | 130.2 | **152.4** | 170.2 |
| Crawl | 8.7 | 26.4 | **45.3** | 58.3 |
| Crouch | 47.1 | 62.8 | **64.2** | 77.6 |
| Lay down | 63.9 | 89.1 | **101.4** | 110.7 |
| Sit down | 40.8 | 65.7 | **77.1** | 84.2 |
| Getting up | 49.7 | 58.3 | **67.8** | 80.4 |
| Aiming | 33.6 | 39.2 | **39.8** | 42.3 |
| Walk to High Foot Kick | 67.4 | 89.7 | **95.1** | 108.1 |
| Arm Push / Stretched | 49.3 | 55.8 | **57.4** | 62.5 |
| Low- to High-Ground Walk | 38.8 | 29.4 | **44.7** | 54.9 |
| High- to Low-Ground Walk | 39.1 | 33.9 | **52.8** | 58.2 |

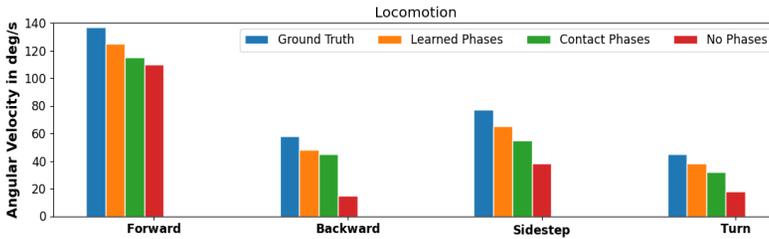

Fig. 5. Angular joint updates for different locomotion styles synthesized by the network with/without using phase features.

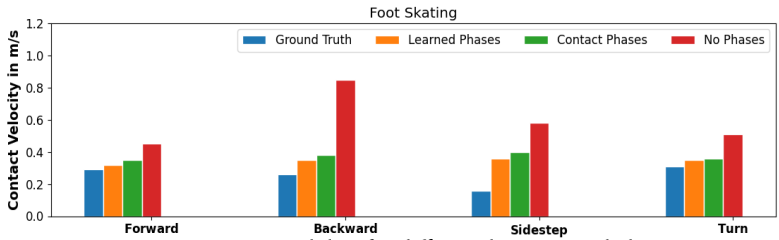

Fig. 6. Foot contact stability for different locomotion behaviors.

## 5.2 Impact of Bi-Directional Control

In this experiment, the precision of reaching the target posture is examined. The precision is measured by the rotational error and positional error after the in-betweening time when the character should be at the targets location. Results for different transition types are shown in Table 2. Without bi-directional control, the character can not correctly align to the target posture. For walking it can be observed that the character uses different stepping patterns. A similar behavior is observed for obstacle locomotion dependent on the given target posture.

## 5.3 Reconstruction Quality and Robustness

We evaluate how well the model is able to reconstruct a large variety of different movement learned from the data. In this experiment, we report the average L2 distances of global quaternions (L2Q) and global positions (L2P) between the ground truth and generated in-betweening motions. We compare our framework to RTN [Harvey et al. 2020] and a basic interpolation in which the joint positions are linearly blended and quaternions spherically interpolated between the keyframes along the in-betweening time. Results are shown in Table 3 on different in-betweening duration times. The proposed framework outperforms the baseline method and RTN, in which it achieves an overall better reconstruction accuracy with more natural movements and less artefacts. In particular, the framework is evaluated on robustness to variable lengths of transitions: The model was trained on transitions of random lengths up to 2 seconds (60 frames). Although trained only on max 60 frames, the framework can manage longer transitions in a good quality (see supplementary video). This indicates that the system can extrapolate for large time deltas by making use of the predicted frequency/phase updates; this forces the motion to keep progressing.

Table 2. The average positional error (in cm) and rotational error (in degree) produced by the model with/without bi-directional control for different tasks in 30 frames transitions windows, following reference control signals in the testing set. The error is measured after the in-betweening time on all joints correspondingly.

| Technique | Walk | | Sit | | Crawl | | Obstacle | |
|---|---|---|---|---|---|---|---|---|
| | cm | deg | cm | deg | cm | deg | cm | deg |
| Non Bi-Directional | 21.59 | 18.94 | 8.66 | 12.98 | 26.69 | 29.64 | 33.89 | 36.42 |
| Bi-Directional | **9.61** | **11.23** | **4.73** | **8.40** | **12.05** | **15.01** | **15.34** | **14.72** |

In contrast, previous methods fail to interpolate long transitions. RTN [Harvey et al. 2020] fails in longer time horizons, especially if the start and end frame are not within the same phase of a walking cycle (i.e. left leg is behind right leg and vice versa); the motion then tends to drift to the target without changing the pose. RTN also struggles when non-common gaits between keyframes have to be performed where the start and end frame are not within the same phase of a walking motion, i.e. one key frame has the left foot planted and the other frame has the right foot planted [Harvey et al. 2020]. Tang et al. [2022] only show generated transition lengths with a maximum of 45 frames on the same dataset. Finally, previous systems tend to fail synthesizing motions where the training samples are sparse, such as side stepping and backward walking.

*Stress Test.* In Figure 7, an extreme forwarding case is shown, where the target frame is 10 meters away from the starting frame and the longest distance for such transition in the training set is merely 5.79 meters [Tang et al. 2022]. We visually compare the results generated without phases, using contact phases [Starke et al. 2020] and our method using learned phases. Without the phase, the synthesized poses accelerate unnaturally and produce an overestimated running motion with significant foot skating. Using contact phases, the motion produces a more stable running motion but drifts towards the goal during the transition to kneeling down. Lastly, using learned phase variables as input generates the most natural motions under this extreme case by launching a jumping action to reach the target in minimal time, producing a full-body motion that is more consistent to the required acceleration and deceleration. Another stress test is done to let the character to interpolate two keyframes of more than 30 meters apart in 10 seconds. Our system can still produce a reasonable running motion, although the motion is a little blocky in the beginning. The readers are referred to the supplementary video for the details.

Table 3. Comparison on the in-betweening stability for multiple transition lengths with different methods.

| Frames | 30 | 45 | 60 | 75 | 90 | 105 | 120 | AVG |
|---|---|---|---|---|---|---|---|---|
| | | | | **L2P** | | | | |
| Interp. | 2.35 | 3.27 | 4.85 | 6.49 | 7.94 | 9.39 | 11.62 | 6.56 |
| RTN | **2.23** | 2.98 | 3.71 | 3.96 | 4.14 | 4.87 | 5.59 | 3.93 |
| Ours | 2.71 | **2.93** | **3.08** | **3.27** | **3.2** | **3.56** | **3.89** | **3.32** |
| | | | | **L2Q** | | | | |
| Interp. | 0.97 | 1.42 | 1.65 | 1.83 | 1.95 | 2.03 | 2.21 | 1.72 |
| RTN | 0.71 | 0.92 | 1.04 | 1.2 | 1.39 | 1.55 | 1.52 | 1.19 |
| Ours | **0.49** | **0.55** | **0.65** | **0.62** | **0.67** | **0.71** | **0.73** | **0.63** |
| | | | | **Foot skate** | | | | |
| Ground Truth | | | | 0.32 | | | | |
| Interp. | 2.07 | 2.24 | 2.45 | 2.73 | 2.91 | 3.04 | 3.22 | 2.66 |
| RTN | **0.42** | **0.45** | 0.49 | 0.51 | 0.6 | 0.68 | 0.73 | 0.55 |
| Ours | 0.43 | **0.45** | **0.48** | **0.46** | **0.46** | **0.48** | **0.46** | **0.46** |
| | | | | **Angular joint updates** | | | | |
| Ground Truth | | | | 63.0 | | | | |
| Interp. | 14.1 | 12.7 | 11.3 | 9.9 | 8.5 | 7.1 | 5.6 | 9.89 |
| RTN | **62.2** | 58.9 | 51.4 | 47.6 | 41.2 | 31.1 | 27.3 | 45.7 |
| Ours | 61.5 | **58.3** | **58.0** | **59.4** | **52.8** | **48.9** | **47.6** | **55.21** |

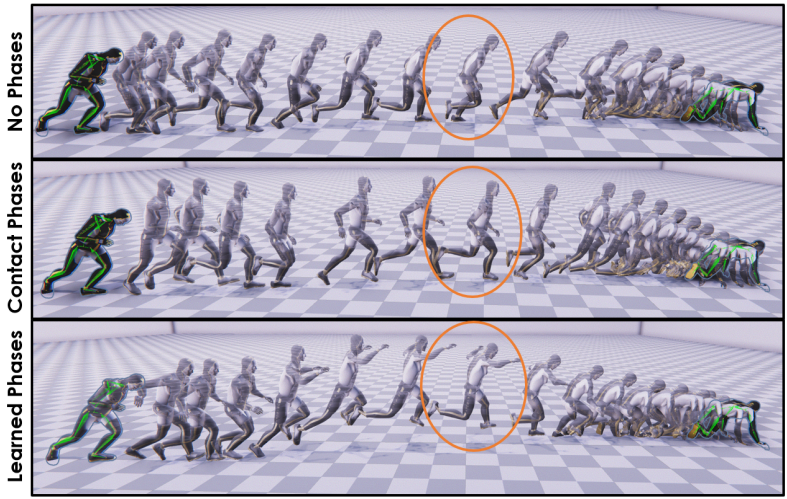

Fig. 7. Example of an extreme forwarding case where the target frame is 10 meters away from the starting frame and has to be reached within 2 seconds. For clarity, only one in four generated frames is shown. For a clearer observation on the motion diversity, one frame from each sequence is marked. The results indicate that the model using learned phases generates the most natural motion transition.

## 5.4 Optional Control Constraints

We evaluate the difference in motion quality depending on what amount of information can be provided by the artist, i.e. root trajectory information or style variables (see Fig. 10).

For the root trajectory, a control parameter $\tau$ is used to blend a future desired trajectory $T^*$ and the model-predicted trajectory $T^+$ to obtain the input trajectory $T$ for the next frame, which can be calculcated as $T = \mathcal{I}(T^*, T^+, \tau)$ where $\mathcal{I}$ is a linear or spherical interpolation for position or rotation features respectively. In Table 4, we measure results on how precisely the character can follow a variety of predefined paths while reaching the target poses, as well as the averaged position and rotation error for each bone at the end of the in-betweening time. We noticed that providing a root trajectory as input can improve the animation quality particularly in terms of foot skating artefacts and stabilizing the motion. With that, it provides additional but optional control channel for the user to produce a larger range of motion variations.

In addition, it can be desired to control the style of the motion to be generated for the in-betweened frames. For example, assuming two upright standing poses as shown in Fig. 8, our method enables authoring which style (e.g. crawling, aiming) the model should generate while reaching the target pose. This can be done by training the model with additional one-hot labels to indicate the desired style during runtime. When trained without action labels, the model produces a normal walking transition between the standing poses (top row) and when using the model with active action labels, a crawling transition can be generated (bottom row).

## 5.5 Time Savings for Prototyping

We also conduct a user study to evaluate the effectiveness of our method. Two professional animators that work on AAA game titles were invited to complete three animation transitions with and without the tool (six tasks in total). Given a starting pose and target pose, the task was to animate a 60, 90, and 120 frame transition at 30Hz. Results in Table 5 show the amount of time spent for manually creating transitions that would be of similar motion quality as generated by the model. This indicates that considerable time savings can be achieved by synthesizing movements in just a few minutes using our framework that would typically require a day when done manually.

Table 4. Impact of controlling the future trajectory through predefined paths on the ground using the authoring tool. The paths with their target frames at each control point are the same as shown in Fig. 10. The in-betweening time between the poses is 2 seconds. The control parameters $\tau$ enables to perform realistic motion transitions along artificial paths, enabling better target accuracy and motion quality.

| Traj. Control $\tau$ | Position Deviation (cm) | | | |
| --- | --- | --- | --- | --- |
| | Circle | Square | Star | Custom |
| 0 (autoregressive) | 11.27 | 14.46 | 15.2 | 11.48 |
| 0.25 | 6.27 | 7.5 | 9.71 | 7.42 |
| 0.5 | 4.65 | 5.46 | 8.47 | 6.96 |
| 0.75 | 3.89 | 4.67 | 8.1 | 7.37 |
| 1.0 | 3.19 | 4.1 | 7.59 | 6.66 |
| **Foot Skate** | | | | |
| 0 (autoregressive) | 0.432 | 0.62 | 0.63 | 0.531 |
| 0.25 | 0.428 | 0.60 | 0.614 | 0.494 |
| 0.5 | 0.416 | 0.577 | 0.571 | 0.48 |
| 0.75 | 0.40 | 0.549 | 0.552 | **0.472** |
| 1.0 | **0.388** | **0.538** | **0.53** | 0.473 |
| **PE (cm/bone)** $\mid$ **RE (deg/bone)** | | | | |
| 0 (autoregressive) | 1.93 $\mid$ 1.21 | 2.34 $\mid$ 1.30 | 2.48 $\mid$ 1.16 | 2.29 $\mid$ 1.09 |
| 0.25 | 1.20 $\mid$ 0.98 | 1.42 $\mid$ 1.08 | 1.15 $\mid$ **0.88** | 1.34 $\mid$ 0.86 |
| 0.5 | 1.02 $\mid$ 0.81 | 1.10 $\mid$ **1.01** | 0.91 $\mid$ 0.92 | 1.12 $\mid$ 0.82 |
| 0.75 | 0.87 $\mid$ **0.79** | 0.98 $\mid$ 1.06 | 0.82 $\mid$ 0.96 | 1.09 $\mid$ 0.78 |
| 1.0 | **0.78** $\mid$ 0.81 | **0.86** $\mid$ 1.05 | **0.72** $\mid$ 0.94 | **0.98** $\mid$ **0.77** |
| **Angular joint updates** | | | | |
| 0 (autoregressive) | **73.9** | 77.6 | 70.7 | 71.1 |
| 0.25 | 73.1 | 77.0 | 72.1 | 72.6 |
| 0.5 | 72.1 | 77.2 | 74.1 | 74.0 |
| 0.75 | 72.4 | 78.3 | 75.9 | **75.2** |
| 1.0 | 72.8 | **79.0** | **76.2** | 74.9 |

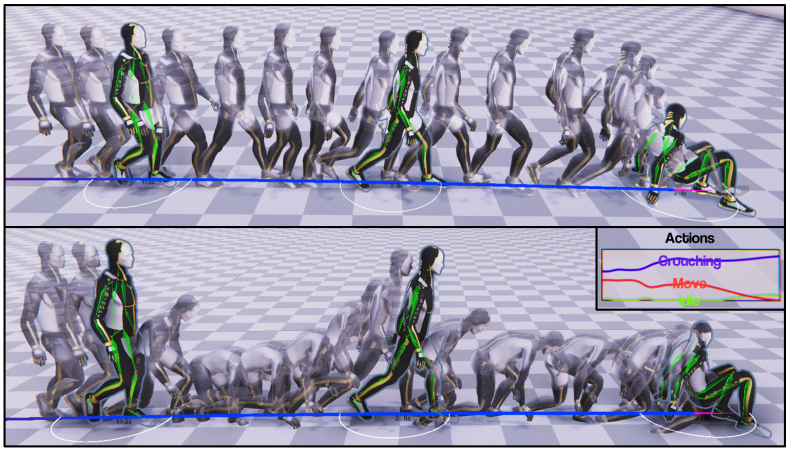

Fig. 8. Motions produced by the model with (bottom) and without (top) action labels in 60 frames transition windows. The blue path indicates that the character should crawl between the target frames.

Table 5. The average time taken by two professional animators to complete three animation transitions in our user study.

| Transition | Frames | without Tool | with Tool |
|---|---|---|---|
| Running cycle | 60 | 5.24h | 6.5min |
| Lay-to-Walk | 90 | 7.61h | 9.7min |
| Walk-to-Crawl | 120 | 9.5h | 7.9min |

Table 6. Qualitative results on quadruped dataset [Zhang et al. 2018]. The models are trained with a maximum transition length of 60 frames (1 second).

| | L2P | | | L2Q | | |
|---|---|---|---|---|---|---|
| **Frames** | **30** | **60** | **90** | **30** | **60** | **90** |
| Interp. | 0.70 | 1.26 | 1.96 | 0.56 | 0.72 | 0.84 |
| Transformer | **0.35** | 0.63 | 1.27 | **0.31** | 0.46 | 0.66 |
| Ours | 0.39 | **0.51** | **0.87** | 0.36 | **0.42** | **0.55** |

## 5.6 In-betweening on Quadrupeds

Finally, we demonstrate our system on a quadruped character where different locomotion modes such as walk, trot, and pace can be performed well. To do so, we use the same data as in [Zhang et al. 2018] which contains mainly locomotion (77.67%) behaviors. By training a different Periodic Autoencoder and feeding the extracted learned phases into a new motion generator the system is able to reproduce those different locomotion modes between the target frames and synthesize motion transitions at 60Hz. We compare our method with the interpolation baseline and Transformer [Qin et al. 2022]. Quantitative results are presented in Table 6 and qualitative results can be seen in the accompanied video.

## 6 LIMITATIONS AND FUTURE WORK

Similar to other data-driven frameworks, the generated results are limited by the training data and cannot synthesize motions that are too different from the training data. The framework itself may not perfectly accurate achieve the target pose due to the network compression. However, an additional inverse kinematics logic can be implemented to reduce the offset error and solve the minor difference for end-effector locations when the character is near the target state. Another research goal would be to involve interactions with the environment during transition by including environment sensors and interaction sensors to the system.

Although the in-betweening duration provided to the model can span longer deltas than those in previous works, this time value should still be set to a reasonable value and needs to be provided by the user during inference. Learning such time distributions in a stochastic manner could be one future direction to explore. Moreover, making use of the recent advance in diffusion models [Tevet et al. 2022] could allow synthesizing a variation of movements from the same set of keyframes with subtle human movements that can increase the realism of the generated motions. Finally, it would be interesting to look into a motion in-betweening scheme that can manage multi-character interactions [Shum et al. 2008, 2010; Zhang et al. 2023].

## 7 CONCLUSION

We propose a novel learning-based framework for in-betweening character movements from motion capture data. By utilizing phase variables, our autoregressive time series model can generate a wide variety of plausible motion transitions with variable lengths. The system is robust to unseen control signals and especially can generalize to longer transitions exceeding the maximum transition length seen during training with better quality than existing methods. In addition, we propose an authoring tool that provides optional control channels for the user on the task where the style of the motion can be controlled. The system enables faster prototyping workflows for animators.

## ACKNOWLEDGMENTS

We want to thank Ubisoft La Forge for their work on capturing the LAFAN1 animation dataset with its human character model. We further want to thank Matteo Loddo for skinning the quadruped character model and Chris Warnock for his various support throughout this project. Finally, we also wish to thank the anonymous reviewers for their constructive comments.

## A APPENDIX

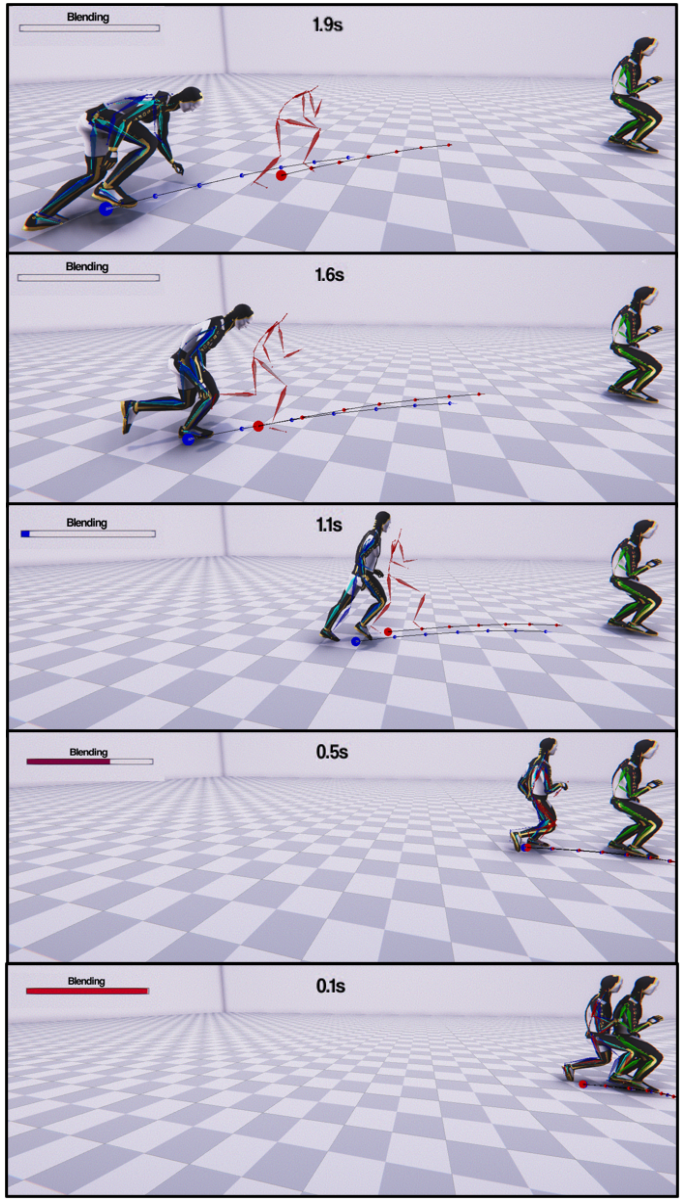

Fig. 9. The bi-directional control scheme enables to reach the desired target pose. The root update, future trajectory and character pose are predicted both relative to the character root (blue) as well as to the coordinate frame of the target (red), and interpolated with a blending parameter (top left) before given as the input for the next frame. The actual character pose (in cyan) is naturally moving towards the target pose (in green) conditioned by the in-betweening time (top middle).

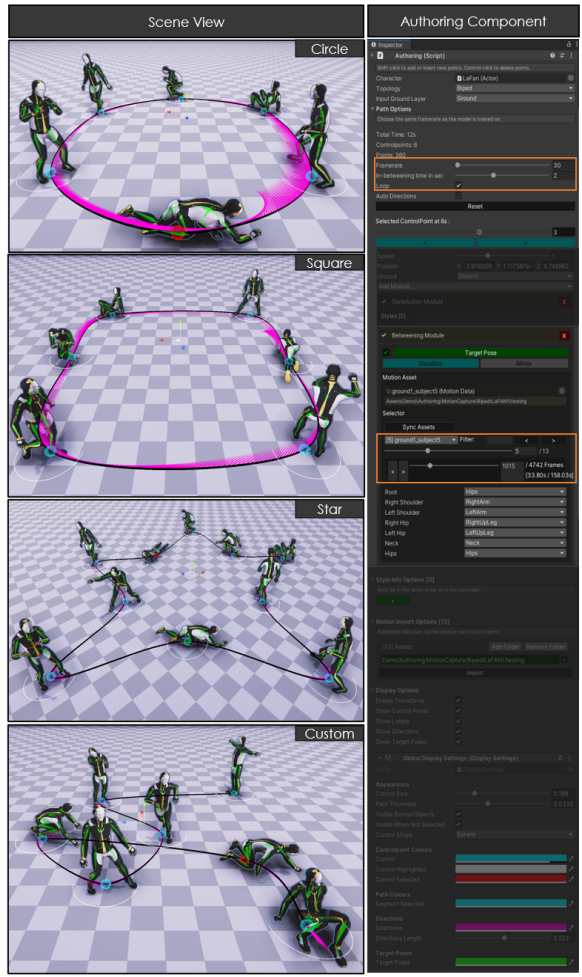

Fig. 10.  Sequences of keyframes along artificial paths (left) created with the authoring tool (right).

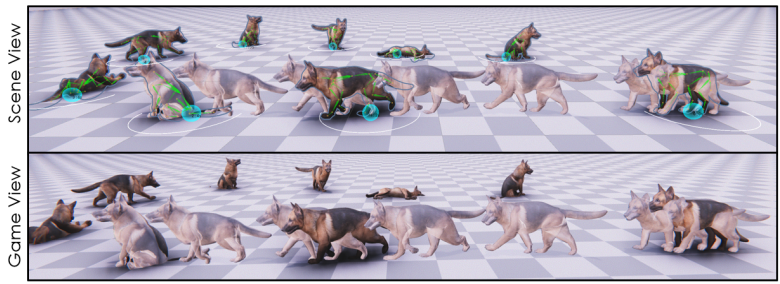

Fig. 11.  Generated in-between motion sequences of a quadruped character between target frames using the proposed framework and authoring tool.